\documentstyle[aps
,epsfig%
,preprint%
]{revtex}
\newcommand{\nn}{\nonumber \\}

\newcommand{\Tr}{\mbox{Tr}}

\begin{document}
\draft
\preprint{Guchi-TP-010}
\date{\today%
}
\title{
Noncommutative gravity in three dimensions
coupled to spinning sources
}
\author{
Kiyoshi~Shiraishi${}^{1}$%
\thanks{Email address: {\tt shiraish@sci.yamaguchi-u.ac.jp}},
Kenji~Sakamoto${}^{2}$%
\thanks{Email address: {\tt b1795@sty.cc.yamaguchi-u.ac.jp}}
and
Nahomi~Kan${}^{2}$%
\thanks{Email address: {\tt b1834@sty.cc.yamaguchi-u.ac.jp}}
}
\address{
${}^{1}$
Faculty of Science, Yamaguchi University,
Yoshida, Yamaguchi-shi, Yamaguchi 753-8512, Japan \\
${}^{2}$Graduate School of Science and Engineering, Yamaguchi University, 
Yoshida, Yamaguchi-shi, Yamaguchi 753-8512, Japan 
}
\maketitle
\begin{abstract}
Noncommutative gravity in three dimensions 
with vanishing cosmological constant is examined. 
We find a solution which describes a spacetime in the presence of
a torsional source. We estimate the phase shift for
each partial wave of a scalar field in the spacetime
by the Born approximation.
\end{abstract}



\section{Introduction}

We expect that noncommutative field theory
\cite{DN} is worth studying,
because the fundamental minimal length scale 
in the theory may avoid singularities and infinities which arise in usual
field theories.
Several attempts to formulate models of gravity on noncomutative
spaces have recently been developed\cite{Nair,Cham,BCGSS,CKMZ,CCKMSZ,KSproc}.
For three dimensional models, a treatment based on the Chern-Simons gauge theory
has been studied in\cite{BCGSS,CKMZ,KSproc}.

In the present paper, we consider three dimensional noncommutative
gravity with no cosmological term. We show an exact solution with
a localized source
in the theory. The size of the point-like structure is controlled by
the length scale $\sqrt{\theta}$ of the theory.
The investigation of scattering by the localized object reveals
how we can ``see'' the noncommutative space by a wave probe.

The plan of the present paper is as follows:
In Sec.~\ref{sec:2} we give a brief description of noncommutative space
for self-containedness as well as for later use of notational convensions.
In Sec.~\ref{sec:3} we review noncommutative gravity
in three dimensions with vanishing cosmological constant.
Sec.~\ref{sec:4} describes spacetime in the presence of torsional sources.
The scattering of a scalar wave is studied in Sec.~\ref{sec:5}.
Finally, Sec.~\ref{sec:f} contains conclusion and discussion.

\section{noncommutative plane}
\label{sec:2}

In this section, we review two dimensional noncommutative space.
For more details, please consult a comprehensive review by 
Douglas and Nekrasov\cite{DN}.
 
Consider noncommutative coordinates, for example,
\begin{equation}
\left[\,x, y\,\right]=i\,\theta\, ,
\end{equation}
where $\theta$ is a real, positive constant.
Then the ``uncertainty'' lies between $x$ and $y$, namely,
\begin{equation}
\Delta x\Delta y\ge \theta\, ,
\end{equation}
(where a numerical factor has omitted).
This implies existence of a minimal length scale $\approx\sqrt{\theta}$.
If complex combinations of the coordinates,
$z=x+iy$ and $\bar{z}=x-iy$, are introduced, they satisfy
\begin{equation}
\left[\,z, \bar{z}\,\right]=2\theta\, .
\end{equation}

There are two different representations to describe the noncommutativity;
the commutative coordinate formalism with the star product and the Fock space
(operator) formalism (see \cite{DN} for details). In this paper, we
simply use expressions in two formaisms identically; 
thus we suppress the star product symbol 
in this paper.%
\footnote{
The differentiation is expressed as
$\partial_zf=[f,\bar{z}]/(2\theta)$ and
$\partial_{\bar{z}}f=[z,f]/(2\theta)$.
At the same time, an integration $\int d^2x f(z,\bar{z})$ in the
noncommutative theory means $2\pi\theta\sum_n f_{nn}$ for $f(z,\bar{z})=\sum_{mn}
f_{mn}|m\rangle\langle n|$.\cite{DN}}
 For example, we write the equalities as
\begin{equation}
1=\sum_{n=0}^{\infty}|n\rangle\langle n|\, ,\qquad
z=\sqrt{2\theta}\sum_{n=0}^{\infty}\sqrt{n+1}|n\rangle\langle n+1|\, ,\qquad
\bar{z}=\sqrt{2\theta}\sum_{n=0}^{\infty}\sqrt{n+1}|n+1\rangle\langle n|\, ,
\end{equation}
where ket and bra satisfy $z|0\rangle=0$, 
$z|n\rangle=\sqrt{2\theta}\sqrt{n}|n-1\rangle$ $(n=1, 2,\ldots)$,
$\bar{z}|n\rangle=\sqrt{2\theta}\sqrt{n+1}|n+1\rangle$ 
$(n=0, 1, 2,\ldots)$, and so on.
Another example is
\begin{equation}
|n\rangle\langle n+\ell|=2 (-1)^n\sqrt{\frac{n!}{(n+\ell)!}}
\left(\frac{2r^2}{\theta}\right)^{\ell/2}
 L_n^{\ell}(2r^2/\theta) e^{-r^2/\theta} e^{i\ell\varphi}\, ,
\end{equation}
where $r^2=x^2+y^2$, $z=r e^{i\varphi}$ and $L_n^{\ell}(x)$ is the
Lagurre polynomial.
In particular, $|0\rangle\langle 0|=2 e^{-r^2/\theta}$.

For later use, we {\it define}%
\footnote{Note that here ${\tt \frac{1}{z}}$ is defined in operator
formalism and this differs from $z^{-1}$ in the usual star product 
formalism unless $\theta=0$.} 
the inverses of $z$ and $\bar{z}$ as
\begin{eqnarray}
{\tt
\frac{1}{z}}&\equiv&\frac{1}{\sqrt{2\theta}}\sum_{n=0}^{\infty}\frac{1}{\sqrt{n+1}}
|n+1\rangle\langle n|=\frac{1}{r}(1-e^{-r^2/\theta}) e^{-i\varphi}\, ,\\
{\tt \frac{1}{\bar{z}}}&\equiv&\frac{1}{\sqrt{2\theta}}\sum_{n=0}^{\infty}
\frac{1}{\sqrt{n+1}}|n\rangle\langle n+1|=\frac{1}{r}(1-e^{-r^2/\theta})
e^{i\varphi}\, .
\end{eqnarray}

This definition leads to $z{\tt \frac{1}{z}}={\tt
\frac{1}{\bar{z}}}\bar{z}=1$, however, one can see
\begin{equation}
{\tt \frac{1}{z}} z=\bar{z}{\tt \frac{1}{\bar{z}}}=1-|0\rangle\langle 0|\, .
\end{equation}

Thus the derivative of ${\tt \frac{1}{z}}$ and ${\tt \frac{1}{\bar{z}}}$ 
turns out to be
\begin{equation}
\partial_{\bar{z}}{\tt \frac{1}{z}}=
\frac{1}{2\theta}\left[z,{\tt \frac{1}{z}}\right]=
\frac{1}{2\theta}|0\rangle\langle 0|=
\frac{1}{\theta}e^{-r^2/\theta}\, ,\quad 
\partial_{z}{\tt \frac{1}{\bar{z}}}=
\frac{1}{2\theta}\left[{\tt \frac{1}{\bar{z}}}, \bar{z}\right]=
\frac{1}{2\theta}|0\rangle\langle 0|=
\frac{1}{\theta}e^{-r^2/\theta}\, .
\end{equation}

Interestingly enough, in the commutative limit, we find
\begin{equation}
\frac{1}{\theta}e^{-r^2/\theta}\stackrel{\theta\rightarrow 0}{\longrightarrow}
\pi\, \delta(x)\delta(y)\, .
\end{equation}

\section{Three dimensional noncommutative gravity}
\label{sec:3}

Throughout this paper, we concentrate our attention on noncommutative gravity 
in three dimensions.%
\footnote{See references\cite{Nair,Cham} for an extension to the other dimensions.}
Three dimensional Chern-Simons noncommutative gravity has recently been studied
by Ba\~nados {\it et al.}\cite{BCGSS} and 
more recently by Cacciatori {\it et al.}\cite{CKMZ}.
We would like to study noncommutative gravity in three dimensions
without a cosmological constant\cite{KSproc}. Further, we consider the case
that spatial coordinates are mutually noncommutative.
The signature is taken to be Euclidean, and the coordinates are 
denoted as
\begin{equation}
x^1=x\, ,\quad
x^2=y\, ,\quad
x^3=\tau\, ,\qquad
{\rm where}\quad \left[\,x, y\,\right]=i\,\theta.
\end{equation}
We define a matrix-valued dreibein one-form and a connection one-form as
\begin{equation}
e=e^a J_a+e^4 i\, ,\quad \omega=\omega^a J_a+\omega^4 i\, ,
\end{equation}
where $e^a=e^a_{\mu}dx^{\mu}$, $\omega^a=\omega^a_{\mu}dx^{\mu}$ and
$J_a$ is expressed by using the Pauli matrix:
\begin{equation}
J_1=\frac{i}{2}\sigma_1,\quad
J_2=-\frac{i}{2}\sigma_2,\quad
J_3=\frac{i}{2}\sigma_3\, .
\end{equation}
A matrix-valued torsion two-form and a curvature two-form
are given by:\cite{BCGSS}
\begin{equation}
{\cal T}=de+\omega\wedge e+e\wedge\omega\, ,
\qquad {\cal R}=d\omega+\omega\wedge\omega\, .
\label{eq:TR}
\end{equation}
In our notation, for example, a vacuum solution of ``Einstein equation''
satisfies
\begin{equation}
{\cal R}={\cal T}=0\, .
\end{equation}

\section{A solution with a torsional source}
\label{sec:4}

Let us examine the following form of the dreibein:
\begin{equation}
e=\frac{i}{2}\left\{d\tau+\frac{S}{2i}
\left({\tt \frac{1}{z}}dz-{\tt \frac{1}{\bar{z}}}
d\bar{z}\right)\right\}
\left(
\begin{array}{lr}
1 & 0 \\
0 & -1
\end{array}
\right)+\frac{i}{2}
\left(
\begin{array}{cc}
0 & dz \\
d\bar{z} & 0
\end{array}
\right)\, ,
\end{equation}
where $S$ is a constant
and $\omega=0$.
Note that $e^4=\omega^4=0$.

Then using (\ref{eq:TR}) we obtain
\begin{equation}
{\cal T}=\frac{S}{4\,\theta}\,|0\rangle\langle 0|\,d\bar{z}\wedge dz
\left(
\begin{array}{rr}
1 & 0 \\
0 & -1
\end{array}
\right)\, ,\qquad
{\cal R}=0\, .
\end{equation}

In the commutative limit, this corresponds to the spinning solution
obtained by Deser, Jackiw and `t Hooft \cite{DJT} with spin $S$ and vanishing mass.
For a finite $\theta$, the torsional source has a finite extension.
Note that $\Tr~{\cal T}=0$ for any value of $\theta$.

This one-body solution can be generalized to the $N$-body
solution for sources located at ${\tt z_a}$, each carrying spin $S_a$,
$a=1,\dots,N$. One finds the solution as
\begin{equation}
e=\frac{i}{2}\left\{d\tau+\sum_{a=1}^N\frac{S_a}{2i}
\left({\tt \frac{1}{z-z_a}}dz-{\tt \frac{1}{\bar{z}-\bar{z}_a}}
d\bar{z}\right)\right\}
\left(
\begin{array}{rr}
1 & 0 \\
0 & -1
\end{array}
\right)+\frac{i}{2}
\left(
\begin{array}{cc}
0 & dz \\
d\bar{z} & 0
\end{array}
\right)\, ,
\end{equation}
where  
${\tt \frac{1}{z-z_a}}$ and ${\tt \frac{1}{\bar{z}-\bar{z}_a}}$ are defined by
\begin{equation}
{\tt \frac{1}{z-z_a}}=\sum_{m=0}^{\infty} {\tt z_a}^m\left({\tt
\frac{1}{z}}\right)^{m+1}\, ,
\qquad
{\tt \frac{1}{\bar{z}-\bar{z}_a}}=\sum_{m=0}^{\infty}
{\tt \bar{z}_a}^m\left({\tt \frac{1}{\bar{z}}}\right)^{m+1}\, .
\end{equation}

\section{Wave equation}
\label{sec:5}

Now we can write down a wave equation for a scalar field around
an isolated torsional source. One of the possible expression for
a scalar $\Phi$ is
\begin{equation}
\left(\partial_{\mu}e^{\mu}_ae^{\nu}_a\partial_{\nu}-m^2\right)
\Phi=0\, ,
\label{eq:we0}
\end{equation}
where $e^{\mu}_a$ is the inverse of $e_{\mu}^a$.
Here the determinant of the dreibein has been omitted,
since it is unity for the solution in Sec.~\ref{sec:4}.
Substituting the dreibein of the one-body solution in Sec.~\ref{sec:4},
 we find that the equation (\ref{eq:we0}) reduces to
\begin{equation}
\left\{-\partial_t^2+2\left(D_zD_{\bar{z}}+D_{\bar{z}}D_z\right)-m^2\right\}
\Phi=0\, ,
\label{eq:we}
\end{equation}
where we have changed the signature into the Lorentzian one and used the
notation
\begin{equation}
D_zf\equiv \partial_zf-\frac{S}{2i}{\tt \frac{1}{z}}\partial_tf\, ,\qquad
D_{\bar{z}}f\equiv
\partial_{\bar{z}}f+\frac{S}{2i}{\tt \frac{1}{\bar{z}}}\partial_tf\, .
\label{ld}
\end{equation}
The constant $m$ denotes the mass of the scalar particle.

If we choose the following ``monochromatic'' wave with energy $E$
as the wave function $\Phi$:
\begin{equation}
\Phi(x,y,t)=\phi(z,\bar{z}) e^{-iEt}\, ,
\end{equation}
then the stationary function satisfies:
\begin{equation}
\left\{2\left(\tilde{D}_z\tilde{D}_{\bar{z}}+
\tilde{D}_{\bar{z}}\tilde{D}_z\right)+k^2\right\}
\phi(z,\bar{z})=0\, ,
\label{eq:21}
\end{equation}
where
\begin{equation}
\tilde{D}_zf=\partial_zf+\frac{SE}{2}{\tt \frac{1}{z}}f\, ,\qquad
\tilde{D}_{\bar{z}}f=
\partial_{\bar{z}}f-\frac{SE}{2}{\tt \frac{1}{\bar{z}}}f\, ,
\end{equation}
and the wave number $k$ is defined as $k=\sqrt{E^2-m^2}$.

In the case with $S=0$, or Minkowski space, the
scalar wave equation becomes
\begin{equation}
\left\{2\left(\partial_z\partial_{\bar{z}}+
\partial_{\bar{z}}\partial_z\right)+k^2\right\}
\chi_{\ell}=0\, ,
\label{eqB}
\end{equation}
and its regular solution $\chi_{\ell}$ is expressed by using
the Bessel function as
\begin{eqnarray}
\chi_{\ell}&=&J_{\ell}(kr) e^{i\ell\varphi}\nn
&=&
\sum_{n=0}^{\infty}\sqrt{\frac{n!}{(n+\ell)!}}
\left(\frac{k^2\theta}{2}\right)^{\ell/2}
L_n^{\ell}(k^2\theta/2) e^{-k^2\theta/4}|n\rangle\langle n+\ell|
\qquad (\ell=0, 1, 2,\ldots)\, ,
\end{eqnarray}
and its asymptotic behavior is
\begin{equation}
\chi_{\ell}
\stackrel{r\rightarrow\infty}{\longrightarrow}
\sqrt{\frac{2}{\pi kr}}
\cos\left(kr-\frac{\ell\pi}{2}-\frac{\pi}{4}\right) e^{i\ell\varphi}\, .
\label{as0}
\end{equation}

Returning to the $SE>0$ case, Eq.~(\ref{eq:21}) can be rewritten as
\begin{equation}
\left\{2\left(\tilde{D}_z\tilde{D}_{\bar{z}}+
\tilde{D}_{\bar{z}}\tilde{D}_z\right)+k^2\right\}
\phi_{\ell}=
2\left(\partial_z\partial_{\bar{z}}+
\partial_{\bar{z}}\partial_z\right)\phi_{\ell}-
U\phi_{\ell}+
k^2\phi_{\ell}
=0\, ,
\end{equation}
where
\begin{equation}
U\phi_{\ell}=2\alpha\left({\tt \frac{1}{\bar{z}}}\partial_z\phi_{\ell}-
{\tt \frac{1}{z}}\partial_{\bar{z}}\phi_{\ell}\right)+
\frac{\alpha^2}{2}\left({\tt \frac{1}{z}{\tt \frac{1}{\bar{z}}}}+
{\tt \frac{1}{\bar{z}}}{\tt
\frac{1}{z}}\right)\phi_{\ell}\, ,
\end{equation}
with $\alpha\equiv SE$. In this paper, we treat the case with
$0<\alpha\ll1$ for simplicity.

Here we assume that $\phi_{\ell}$ takes the similar form to $\chi_{\ell}$:
\begin{equation}
\phi_{\ell}=\sum_{n=0}^{\infty}C_n|n\rangle\langle n+\ell|\, ,
\end{equation}
where $C_n$ depends on $n$, $\ell$, and $k^2\theta$.
This solution may have the following asymptotics:
\begin{equation}
\phi_{\ell}\stackrel{r\rightarrow\infty}{\longrightarrow}
\sqrt{\frac{2}{\pi kr}}
\cos\left(kr-\frac{\ell\pi}{2}-\frac{\pi}{4}+\delta_{\ell}\right) 
e^{i\ell\varphi}\, ,
\label{as1}
\end{equation}
for $r\gg k^{-1}$ and $r\gg\sqrt{\theta}$. 
We call a constant $\delta_{\ell}$ as a phase shift.
Now assuming the regularity
of the solution at the origin and using asymptotic forms (\ref{as0}) and 
(\ref{as1}), we find
\begin{equation}
\int d^2x \left\{2\chi_{\ell}^{\dagger}\left(\partial_z\partial_{\bar{z}}+
\partial_{\bar{z}}\partial_z\right)\phi_{\ell}-2\phi_{\ell}^{\dagger}
\left(\partial_z\partial_{\bar{z}}+
\partial_{\bar{z}}\partial_z\right)\chi_{\ell}\right\}
=-4\sin\delta_{\ell}\, .
\end{equation}
Therefore the phase shift for a partial wave for $\ell\ge 0$ is
written as
\begin{equation}
\sin\delta_{\ell}=-\frac{1}{4}\int d^2x
\chi_{\ell}^{\dagger}U\phi_{\ell}\, .
\end{equation}

Similarly, since $\chi_{\ell}^{\dagger}$ also is a solution for Eq.~(\ref{eqB}),
we take $\chi_{-\ell}\equiv\chi_{\ell}^{\dagger}$ and
$\phi_{-\ell}\equiv\phi_{\ell}^{\dagger}$ for $\ell\ge 0$. 

When $\alpha$ is sufficiently small, the Born approximation may be valid
to obtain
\begin{equation}
\delta_{\ell}\simeq -\frac{1}{4}\int d^2x
\chi_{\ell}^{\dagger}U\chi_{\ell}\, ,\qquad
\delta_{-\ell}\simeq -\frac{1}{4}\int d^2x
\chi_{\ell}U\chi_{\ell}^{\dagger}\qquad
{\rm for~\ell\ge 0}\, .
\end{equation}

The calculation done by the operator formalism leads to%
\footnote{See Appendix for formulas on the Laguerre functions.}
\begin{eqnarray}
& &\frac{1}{2\pi\alpha}\int d^2x
\chi_{\ell}^{\dagger}(2\alpha)\left({\tt \frac{1}{\bar{z}}}\partial_z\chi_{\ell}-
{\tt \frac{1}{z}}\partial_{\bar{z}}\chi_{\ell}\right)\nn
&=&2\left(\frac{k^2\theta}{2}\right)^{\ell} e^{-k^2\theta/2}
\sum_{n=0}^{\infty}\frac{n!}{(n+\ell)!}L_n^{\ell}(k^2\theta/2)
(L_{n+1}^{\ell}(k^2\theta/2)-L_n^{\ell}(k^2\theta/2))\nn
& &+\frac{1}{\ell!}\left(\frac{k^2\theta}{2}\right)^{\ell} e^{-k^2\theta/2}
(L_0^{\ell}(k^2\theta/2))^2\nn
&=&1-2\frac{\gamma(\ell,k^2\theta/2)}{\Gamma(\ell)}+
\frac{1}{\ell!}\left(\frac{k^2\theta}{2}\right)^{\ell}
e^{-k^2\theta/2}\nn
&=&1-2\frac{\gamma(\ell+1,k^2\theta/2)}{\ell !}-
\frac{1}{\ell !}\left(\frac{k^2\theta}{2}\right)^{\ell}
e^{-k^2\theta/2}\qquad (\ell>0)\, ,
\end{eqnarray}
where the incomplete gamma function
\begin{equation}
\gamma(\ell+1,y)=\int_0^y e^{-t}t^{\ell}dt\, ,
\end{equation}
has been used.
In the same manner, one finds
\begin{eqnarray}
& &\frac{1}{2\pi\alpha}\int d^2x
\chi_{0}^{\dagger}(2\alpha)\left({\tt \frac{1}{\bar{z}}}\partial_z\chi_{0}-
{\tt \frac{1}{z}}\partial_{\bar{z}}\chi_{0}\right)\nn
&=&2 e^{-k^2\theta/2}
\sum_{n=0}^{\infty}L_n(k^2\theta/2)
(L_{n+1}(k^2\theta/2)-L_n(k^2\theta/2))+ e^{-k^2\theta/2}
(L_0(k^2\theta/2))^2\nn
&=&-1+e^{-k^2\theta/2}\qquad (\ell=0)\, ,
\end{eqnarray}
where $L_n(x)\equiv L_n^0(x)$, as well as
\begin{eqnarray}
& &\frac{1}{2\pi\alpha}\int d^2x
\chi_{\ell}(2\alpha)\left({\tt \frac{1}{\bar{z}}}\partial_z\chi_{\ell}^{\dagger}-
{\tt \frac{1}{z}}\partial_{\bar{z}}\chi_{\ell}^{\dagger}\right)\nn
&=&2\left(\frac{k^2\theta}{2}\right)^{\ell} e^{-k^2\theta/2}
\sum_{n=1}^{\infty}\frac{n!}{(n+\ell)!}L_n^{\ell}(k^2\theta/2)
(L_{n-1}^{\ell}(k^2\theta/2)-L_n^{\ell}(k^2\theta/2))\nn
& &-\frac{2}{\ell!}\left(\frac{k^2\theta}{2}\right)^{\ell} e^{-k^2\theta/2}
(L_0^{\ell}(k^2\theta/2))^2\nn
&=&-1\qquad (\ell>0)\, .
\end{eqnarray}

Consequently, we have the expression for the phase shift
in the Born approximation up to $O(\alpha)$,
\begin{equation}
\delta_{\ell}\simeq
\left\{
\begin{array}{ll}
-\frac{\pi\alpha}{2}\left\{1-2\frac{\gamma(\ell+1,k^2\theta/2)}{\ell !}-
\frac{1}{\ell!}\left(\frac{k^2\theta}{2}\right)^{\ell}
e^{-k^2\theta/2}\right\} & \quad (\ell>0)\\
\frac{\pi\alpha}{2}(1-e^{-k^2\theta/2}) & \quad (\ell=0)\\
\frac{\pi\alpha}{2} & \quad (\ell<0)
\end{array}
\right.
\end{equation}

In the commutative limit, $\theta\rightarrow 0$, this result reduces to
$\delta_{\ell}=-\pi\alpha/2$ for $\ell>0$ and
$\delta_{\ell}=\pi\alpha/2$ for $\ell<0$,
which coincides with the one derived by using
\[
\int_{0}^{\infty}\frac{\ell}{r}(J_{\ell}(r))^2 dr=\frac{1}{2}
\qquad (\ell>0)\, ,
\]
for the ordinary integration.

In the opposite limit, $\theta\rightarrow\infty$, the phase shift for any $\ell$
becomes the  same value
\begin{equation}
\delta_{\ell}=\frac{\pi\alpha}{2} \quad 
{\rm when} \quad \theta\rightarrow\infty\, .
\end{equation}

In the usual commutative model\cite{GJ}, it is known that the first-order 
Born approximation up to
$O(\alpha)$ gives an exact result for the phase shift (except for
$\ell=0$).


\section{conclusion and discussion}
\label{sec:f}

To summarize, we have found an exact solution for three dimensional
noncommutative gravity and studied
the phase shift of scalar waves in the
spinning background spacetime.
The behavior of the phase shift is much alike Aharonov-Bohm scattering;
especially for $\theta=0$ case, the result of de Sousa Gerbert and Jackiw
\cite{GJ} is reproduced.
When $\theta$ is finite, the wave of sufficiently short wave length
can ``see'' the extension of the torsional source, then the difference in
the phase shift is reduced following the order of the angular momentum
$\ell$.

The asymmetric result on the sign of $\ell$ comes from the choice of the wave
equation. In noncommutative theory, left- and right- covariant dervatives give
different results in general, where the right-covariant derivatives read
\begin{equation}
D_z^Rf\equiv \partial_zf-\partial_tf\frac{S}{2i}{\tt \frac{1}{z}}\, ,\qquad
D_{\bar{z}}^Rf\equiv
\partial_{\bar{z}}f+\partial_tf\frac{S}{2i}{\tt \frac{1}{\bar{z}}}\, ,
\label{eq:DR}
\end{equation}
and  the previous ones (\ref{ld}) are the left-derivative.
In other words, if $\Phi$ is a solution of
the wave equation (\ref{eq:we}), $\Phi^{\dagger}$ is not always a solution.

\begin{figure}[htb]
\centering
\mbox{\epsfbox{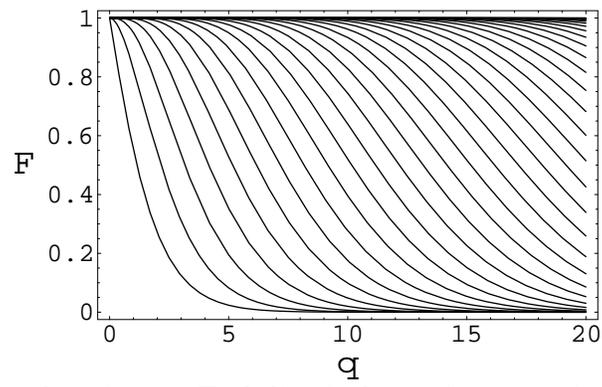}}
\caption{%
$F(p,q)$ is plotted against $q$. Each line is drawn for $p=1, 2, 3,\ldots$, 
from the left to the right.}
\label{fig1}
\end{figure}

The wave equation (\ref{eq:we}) can be expected to be derived from the
lagrangian density
\[-(e_a^{\mu}\partial_{\mu}\Phi)^{\dagger}
(e_a^{\nu}\partial_{\nu}\Phi)-m^2\Phi^{\dagger}\Phi\, .\]
The
lagrangian density
\[\propto -(e_a^{\mu}\partial_{\mu}\Phi)^{\dagger}
(e_a^{\nu}\partial_{\nu}\Phi)-(\partial_{\mu}\Phi e_a^{\mu})^{\dagger}
(\partial_{\nu}\Phi e_a^{\nu})-\cdots\] leads to a wave equation
\begin{equation}
\frac{1}{2}\partial_{\mu}e^{\mu}_ae^{\nu}_a\partial_{\nu}\Phi+
\frac{1}{2}\partial_{\mu}\partial_{\nu}\Phi e^{\nu}_ae^{\mu}_a-m^2
\Phi=0\, ,
\end{equation}
which includes
the left- and right-covariant derivatives 
for our spacetime; this yields a
symmetric result for the phase shift, that is
\begin{equation}
\delta_{\ell}\simeq
\left\{
\begin{array}{cl}
-\frac{\pi\alpha}{2}F(\ell, k^2\theta/2) & \quad (\ell>0)\\
0 & \quad (\ell=0) \\
\frac{\pi\alpha}{2}F(|\ell|, k^2\theta/2) & \quad (\ell<0)
\end{array}
\right.\, ,
\end{equation}
with
\begin{equation}
F(p,q)\equiv
1-\frac{\gamma(p+1,q)}{p\,!}-
\frac{1}{2 p\,!}q^{p}
e^{-q}\, .
\end{equation}
$F(p,q)$ is plotted in Fig.~\ref{fig1}.



In this paper, we have found a noncommutative solution describing
massless spinning sources.
We wonder how we can obtain ``conical'' ({\it i.e.} massive) solutions
in noncommutative gravity.
We also want to know how we take global properties of spacetime into account.
How and when do we have to use a noncommutative torus and sphere?
The choice of global structure of spacetime must be a very important
problem in noncommutative gravity in any dimensions and in any
formulation.


\section*{Acknowledgement}
We would like to thank Y. Cho for reading this manuscript
and for useful comments.

\appendix
\section*{Some useful formulas for the Laguerre functions}

Some useful formulas for the Laguerre functions are listed
as follows\cite{GR}:
\begin{equation}
\frac{e^{xt/(1+t)}}{(1+t)^{\ell+1}}=\sum_{n=0}^{\infty}
(-1)^n L_{n}^{\ell}(x)t^n\, ,
\end{equation}
\begin{equation}
L_{0}^{\alpha}(x)=1\, ,
\end{equation}
\begin{equation}
\sum_{n=0}^{\infty}
\frac{n!}{(n+\ell)!}(-1)^n L_n^{\ell}(x) L_n^{\ell}(y)\, t^n
=\frac{(xyt)^{-\ell/2}}{1+t}e^{t(x+y)/(1+t)}
J_{\ell}(2\sqrt{xyt}/(1+t))\, ,
\end{equation}
\begin{equation}
L_{n}^{\alpha-1}(x)=L_{n}^{\alpha}(x)-L_{n-1}^{\alpha}(x)\, ,
\end{equation}
\begin{equation}
\sum_{n=0}^{\infty}
\frac{n!}{(n+\ell)!} L_n^{\ell}(x)L_{n+1}^{\ell-1}(x)
=\frac{e^{x}}{2
x^{\ell}}\left(1-2\frac{\gamma(\ell,x)}{\Gamma(\ell)}\right)\, ,
\end{equation}
where
\begin{equation}
\gamma(\ell,y)=\int_0^y e^{-t}t^{\ell-1}dt\, ,
\end{equation}
\begin{equation}
\sum_{n=0}^{\infty}
\frac{n!}{(n+\ell)!} L_n^{\ell-1}(x)L_n^{\ell}(x)=\frac{e^{x}}{2 x^{\ell}}\, .
\end{equation}


\end{document}